\pdfoutput=1
\documentclass[11pt,a4paper]{article}

\usepackage[T1]{fontenc}
\usepackage[utf8]{inputenc}
\usepackage{lmodern}
\usepackage[margin=2.75cm,includefoot]{geometry}
\usepackage{microtype}
\usepackage{amsmath,amssymb}
\usepackage{booktabs}
\usepackage{array}
\usepackage{multirow}
\usepackage{placeins}
\usepackage{longtable}
\usepackage{graphicx}
\usepackage{caption}
\usepackage{subcaption}
\usepackage{pgfplots}
\usepackage{xcolor}
\usepackage[numbers,sort&compress]{natbib}
\usepackage[hidelinks]{hyperref}
\usepackage{url}
\hypersetup{pdftitle={Timing Sensitivity in Actuated Traffic Signal Control: A simulation study on one urban network},pdfauthor={Nitai Aharoni}}

\pgfplotsset{compat=1.18}

\definecolor{cFixed}{HTML}{000000}
\definecolor{cBound}{HTML}{8C8C8C}
\definecolor{cRand}{HTML}{8C8C8C}
\definecolor{cGreen}{HTML}{5A5A5A}
\definecolor{cAct}{HTML}{8C2D04}
\definecolor{cCap}{HTML}{D95F02}
\definecolor{cTrunc}{HTML}{E8A33D}
\definecolor{cLQ}{HTML}{0B3D91}
\definecolor{cMP}{HTML}{3690C0}
\definecolor{cMPC}{HTML}{7FB2D6}
\definecolor{cBusy}{HTML}{8C2D04}
\definecolor{cQuiet}{HTML}{0B3D91}
\definecolor{cAny}{HTML}{8C8C8C}
\definecolor{cL1}{HTML}{C8C8C8}
\definecolor{cL2}{HTML}{949494}
\definecolor{cL3}{HTML}{5A5A5A}
\definecolor{cL4}{HTML}{C1272D}
\definecolor{cSig}{HTML}{08519C}
\definecolor{cUnsig}{HTML}{D95F02}

\newcommand{\strat}[1]{\texttt{#1}}

\title{\bfseries Timing Sensitivity in Actuated Traffic Signal Control\\[6pt]
{\large\normalfont A simulation study on one urban network}}
\author{Nitai Aharoni\thanks{Independent researcher, Tel Aviv, Israel.
Correspondence: \texttt{nitaiaharoni1@gmail.com}}}
\date{September 2026}
\begin{document}
\maketitle

\begin{abstract}
A comparison between adaptive and fixed-time traffic signals can change when
only their timing settings change. We examine this sensitivity in 240 simulation
runs on one OpenStreetMap-derived urban network, using eight configurations,
three arrival rates and ten paired demand seeds. The experiment varies the
maximum green of presence-based actuation, retimes the fixed baseline, and
includes a discharge-based phase-termination rule. The endpoint is finite-window
time spent per offered trip, including entry waiting and penalties for removed
or abandoned trips. At the highest tested load, actuation with a maximum green
of twice the planned split has a mean endpoint 57.4 per cent above the original
fixed plan. Reducing that multiplier to 1.25 puts it 70.7 per cent below the
same plan. Against the best of three tested fixed plans, discharge-based
termination reduces the mean endpoint by 14.8, 15.8 and 18.3 per cent across the
three loads. Both timing sweeps favour their shortest tested setting, so neither
identifies an optimum. This exploratory study uses synthetic demand, negligible
start-up lost time and physical removal of persistently stationary vehicles.
The results support testing timing sensitivity before attributing performance
to a controller family; they do not establish a setting for real-world deployment.
\end{abstract}


\section{Introduction}
\label{sec:intro}

Comparisons of traffic-signal controllers depend on the fixed-time reference
and on settings within each responsive controller. If those settings are left
unchanged as demand increases, a comparison can attribute a timing effect to the
control family. This study asks a specific question: how does the comparison
between presence-based actuation and fixed-time control change when maximum
green and fixed-plan duration are varied on the same traffic?

We examine eight configurations on one urban network: four maximum-green
settings for actuation, three fixed plans, and an actuated variant that also
ends green when no discharge is detected. The experiment holds the network,
journeys and remaining controller parameters constant. Its contribution is a
paired sensitivity analysis, including a comparison with a retimed baseline
and explicit accounting for unfinished demand. It is not a new controller or a
benchmark of all adaptive-control families.

Timing sensitivity and discharge-based termination have established precedents.
Signal-timing guidance treats minimum green, passage time, maximum green and
clearance intervals as distinct choices \citep{urbanik2015signal}.
\citet{denney2008saturated} discuss terminating a phase when downstream
restrictions prevent departures despite continuing demand; phase truncation and
flow-based termination are also studied by
\citet{beaird2006truncation,smaglik2006flowbased}. Those works motivate the
comparison here. We do not infer spillback from a lack of departures alone.

\section{Methods}
\label{sec:methods}

\subsection{Network, traffic and simulation}
\label{sec:simulator}

The fixture is an OpenStreetMap extract of central Tel Aviv spanning
32.06 to 32.08 degrees north and 34.77 to 34.79 degrees east, approximately
4.2\,km\textsuperscript{2}. Its constructed network has 2,025 street segments,
2,749 directed lanes and 160.0\,km of lane length. Signalised nodes within
40\,m, up to six at a time, are clustered into 81 signal plans; 78 of these
have two phases. Lane counts and speed limits missing from the extract are
imputed using road-class medians within it. Turn restrictions are retained.
The geometry is real, but demand and signal timings are synthetic.

The simulator uses the Intelligent Driver Model for car following
\citep{treiber2000congested} and MOBIL for lane changing
\citep{kesting2007mobil}, integrated at 0.1\,s. Leader search extends along the
route across junctions to 200\,m. A standstill rule stops a vehicle below
0.1\,m/s when its proposed acceleration is below
0.15\,m/s\textsuperscript{2}. The fleet is 85 per cent cars, 8 per cent buses
and 7 per cent trucks, with driver parameters sampled once per trip.
Appendix~\ref{app:params} supplies vehicle and model parameters.
Appendix~\ref{app:checks} records the available simulator checks and their
known discrepancies; these are separate from the 240-run timing campaign.

Junction movements follow explicit curved connections. Geometric conflicts
define signal phases; permitted turns yield to conflicting movements. At
unsignalised conflicts, drivers use gap acceptance, with critical gaps declining
with waiting time to a 1.5\,s floor. Vehicles require receiving space before
entering a junction; this check looks through links too short to hold them.
Routes minimise free-flow travel time plus turn penalties and remain fixed.
Lane changes cannot alter the destination or route.

Demand is drawn before each run: exponential inter-arrival times and trips
sampled from the connected network, with origins and destinations weighted by
road class. The trip mix is 55 per cent internal, 20 inbound, 20 outbound and
5 through. Internal origins lie along their starting lane. A journey unable to
enter waits in an entry queue and abandons after 300\,s. A vehicle stationary
on the road for 400\,s is physically removed and counted. These interventions
are part of the simulated system, not observed driver behaviour.

The TypeScript simulation core is shared with a headless experiment runner.
Randomness is explicitly seeded, integration and entity ordering are fixed,
and runs use no wall-clock input. Reproducibility has been checked on one
platform, not across JavaScript engines or hardware.

\subsection{Controller configurations}
\label{sec:strategies}

Fixed-time plans divide a specified total green among phases using
geometry-derived discharge weights. A permitted-only lane contributes a weight
of 0.35; planned phase greens have an 8\,s minimum. Yellow and all-red are
computed from approach speed and movement length and lie within 3 to 6\,s
and 1 to 6\,s, respectively. The fixed settings of 40, 60 and 90\,s in
Table~\ref{tab:design} are \emph{total green}, not cycle lengths. All clearance
intervals are added and are served in full in every configuration. Fixed-time
plans use random cycle offsets and are not optimised for demand.

All responsive configurations use the phase structure and planned splits of
the 60\,s reference. They enforce a 7\,s minimum green, scan every 0.1\,s,
and read detectors refreshed at 1\,Hz. Presence is true if a vehicle is within
30\,m of the stop line or if any vehicle is queued on the approach, without a
distance bound for the latter condition. After minimum green, the presence-based
policy ends green at the first scan without demand or at its maximum green.
It chooses the next phase round-robin, skipping phases without demand.
There is no passage timer that holds green briefly after the last detection.
Responsive junctions start on their first phase rather than with random offsets.
Thus the comparison with fixed-time control changes more than maximum green;
only comparisons within the actuated sweep isolate that parameter.

\begin{table}[htbp]
\centering\small
\caption{The eight configurations. Identifiers match the saved run records.
The truncating configuration retains the default maximum-green multiplier of 2.}
\label{tab:design}
\begin{tabular}{@{}lp{8.4cm}@{}}
\toprule
Identifier & Setting or change \\
\midrule
\strat{fixed-green40} & Fixed plan, 40\,s total green \\
\strat{fixed-time} & Fixed plan, 60\,s total green; reference \\
\strat{fixed-green90} & Fixed plan, 90\,s total green \\
\strat{actuated-max125} & Presence-based actuation, maximum $1.25\times$ planned split \\
\strat{actuated-max150} & Presence-based actuation, maximum $1.5\times$ planned split \\
\strat{actuated} & Presence-based actuation, maximum $2\times$ planned split \\
\strat{actuated-max300} & Presence-based actuation, maximum $3\times$ planned split \\
\strat{actuated-truncating} & Default actuation plus termination after 4\,s without discharge \\
\bottomrule
\end{tabular}
\end{table}

For \strat{actuated-truncating}, the discharge clock is reset at the start of
green. The additional condition is checked only after the 7\,s minimum, so a
phase with no departures ends at that minimum, not after four seconds. The
clock uses sampled stop-line discharge counts. The 4\,s threshold was not
swept; neither were minimum green, passage timing, offsets or phase splits.

\subsection{Campaign and pairing}
\label{sec:design}

Eight configurations, three arrival rates and ten seeds give 240 runs. Arrival
rates are 0.12, 0.20 and 0.24 vehicles per second. Their realised mean offers
in the measurement hour are 431.0, 715.3 and 876.2. Seeds 1 to 10 determine the
journeys, vehicle types and driver parameters; at each load every configuration
receives the same schedule. This common-random-numbers design supports paired
comparisons \citep{glasserman1992some}. It does not make runs at different
loads or configurations additional independent demand samples.

Each run has a 32,400\,s warm-up followed by a 3,600\,s measurement window.
Measurement counters reset at the boundary while vehicles and entry queues
remain. The same duration is used for all configurations. The endpoint therefore
describes performance at a common ten-hour horizon; convergence to equilibrium
has not been established.

This campaign followed inspection of an earlier 720-run comparison and reused
its seeds. Ninety of the 240 scenario-seed cells repeat earlier configurations;
their saved metric records agree. They are counted once here and are not
independent confirmation of the earlier results. Settings and policy choices
were developed within that research process, so this is exploratory sensitivity
analysis, not a pre-registered or held-out evaluation.

\subsection{Endpoint and accounting}
\label{sec:metrics}
\label{sec:tts}

Let $W=[t_0,t_1]$ denote the measurement window. The primary endpoint, in seconds,
is
\begin{equation}
Y=\frac{1}{N_{\mathrm{offered}}}
\int_{t_0}^{t_1}\left[N_{\mathrm{road}}(t)+N_{\mathrm{entry}}(t)
+N_{\mathrm{lost}}(t)\right]dt.
\label{eq:tts}
\end{equation}
Here $N_{\mathrm{lost}}$ counts journeys removed or abandoned \emph{within}
the window, accumulating from the moment of loss until $t_1$. It preserves a
remaining-window time charge after a loss. No-route offers would be excluded
from the charged population; none occur in this campaign. The integral is
approximated at the simulation timestep.

This is time spent, not delay above free flow or mean completed-journey time.
It includes entry waiting, which road-delay metrics omit, and unfinished trips,
which completed-trip averages omit. It also includes the backlog inherited
from warm-up. If $N_0$ is that initial population, and $\Delta A(t)$ and
$\Delta D(t)$ count offers and completions since $t_0$, conservation gives
\begin{equation}
N_{\mathrm{offered}}Y
= N_0(t_1-t_0)+\int_{t_0}^{t_1}\Delta A(t)dt
-\int_{t_0}^{t_1}\Delta D(t)dt.
\label{eq:conservation}
\end{equation}
Although offers are paired, $N_0$ depends on the controller. The comparison
therefore includes the state created during warm-up, not just service of new
arrivals in the final hour.

We report initial and final populations, completions, removals and entry
abandonments alongside $Y$. All 240 records satisfy the corresponding flow
balance to numerical precision. The loss charge changes accounting, not physical
traffic: removing a vehicle can still free road space. It excludes losses
before $t_0$ and all time after $t_1$. We consequently interpret $Y$ as a
finite-window network burden per offer, rather than a loss-free capacity or
traveller-welfare measure.

\subsection{Statistical analysis}
\label{sec:stats}

For each non-reference configuration and load, we analyse the ten seed-paired
differences in $Y$. We report the mean difference, a nominal 95 per cent Student
$t$ interval with nine degrees of freedom, and a two-sided $p$ value.
Holm adjustment \citep{holm1979simple} covers the 21 comparisons against the
60\,s reference. A separate family covers 63 comparisons among the seven
non-reference configurations. These are the campaign's original families;
focusing this manuscript has not reduced them to only the comparisons shown
in the main text. Secondary endpoints are descriptive here.

The Student analysis assumes independent seed-level differences and approximate
normality at this sample size. Intervals are unadjusted, not simultaneous.
The exact sign-flip sensitivity calculation enumerates all $2^{10}$ sign
assignments. It avoids normality but still requires sign symmetry or valid
within-pair exchangeability under the null. Its minimum two-sided $p$ value
is $2/1024=0.001953125$; Holm correction over 21 can retain such results,
whereas correction over 63 cannot retain any ten-seed sign-flip result.

Percent reductions divide the paired mean difference by the comparator's
mean. Figure~\ref{fig:timing} similarly shows ratios of configuration means,
not means of seed-level ratios. Tests and intervals remain on paired differences
in seconds. Selecting the best fixed setting after seeing results makes its
comparison exploratory; family-specific corrections do not account for that
selection or all prior research decisions.

\section{Results}
\label{sec:results}
\label{sec:ceiling}

\subsection{The maximum-green setting changes the comparison}

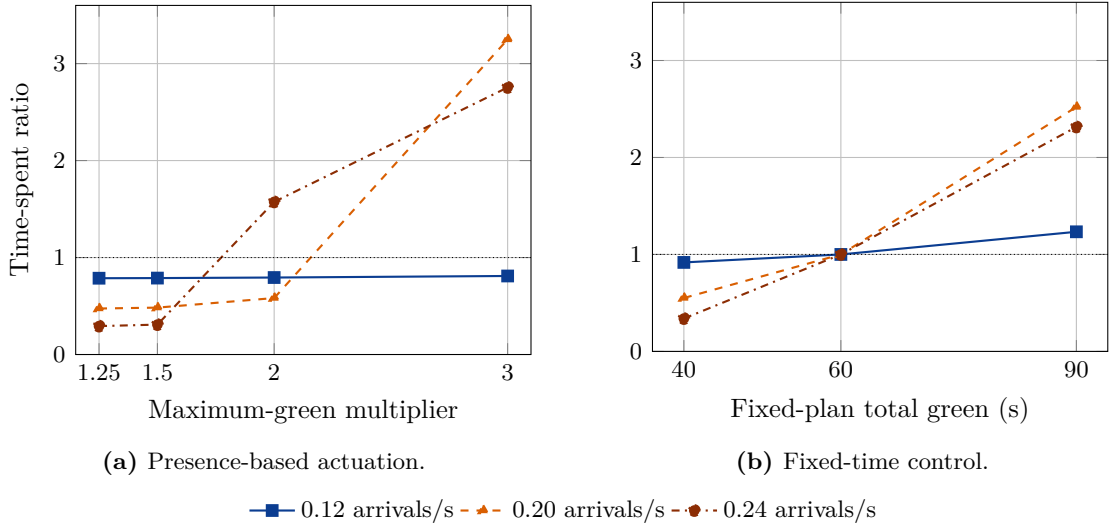
\begin{figure}[htbp]
\centering
\begin{subfigure}[t]{0.49\linewidth}
\centering
\begin{tikzpicture}
\begin{axis}[
width=\linewidth,height=6.2cm,
xmin=1.15,xmax=3.1,xtick={1.25,1.5,2,3},
ymin=0,ymax=3.6,ylabel={Time-spent ratio},
xlabel={Maximum-green multiplier},grid=major,
tick label style={font=\footnotesize},label style={font=\small},
legend to name=timingloads,
legend style={legend columns=3,font=\footnotesize,draw=none},
]
\addplot[cLQ,thick,mark=square*] table[x=setting,y=r12] {data/timing-actuation.dat};
\addlegendentry{0.12 arrivals/s}
\addplot[cCap,thick,dashed,mark=triangle*] table[x=setting,y=r20] {data/timing-actuation.dat};
\addlegendentry{0.20 arrivals/s}
\addplot[cAct,thick,dashdotted,mark=*] table[x=setting,y=r24] {data/timing-actuation.dat};
\addlegendentry{0.24 arrivals/s}
\addplot[black,densely dotted,domain=1.15:3.1,samples=2,forget plot] {1};
\end{axis}
\end{tikzpicture}
\caption{Presence-based actuation.}
\end{subfigure}\hfill
\begin{subfigure}[t]{0.49\linewidth}
\centering
\begin{tikzpicture}
\begin{axis}[
width=\linewidth,height=6.2cm,
xmin=36,xmax=94,xtick={40,60,90},
ymin=0,ymax=3.6,xlabel={Fixed-plan total green (s)},grid=major,
tick label style={font=\footnotesize},label style={font=\small},
]
\addplot[cLQ,thick,mark=square*] table[x=setting,y=r12] {data/timing-fixed.dat};
\addplot[cCap,thick,dashed,mark=triangle*] table[x=setting,y=r20] {data/timing-fixed.dat};
\addplot[cAct,thick,dashdotted,mark=*] table[x=setting,y=r24] {data/timing-fixed.dat};
\addplot[black,densely dotted,domain=36:94,samples=2] {1};
\end{axis}
\end{tikzpicture}
\caption{Fixed-time control.}
\end{subfigure}
\par\smallskip\pgfplotslegendfromname{timingloads}
\caption{Timing sensitivity at three loads. Each point is a ratio of means over
ten paired seeds to the 60\,s fixed plan at the same load. Below the dotted
reference line is better. Connecting lines only guide the eye; the two panels
vary different parameters. Paired intervals in seconds appear in
Appendix~\ref{app:results}.}
\label{fig:timing}
\end{figure}

Figure~\ref{fig:timing} shows the timing response. At 0.24 arrivals per second,
the mean endpoint for default actuation is 2198.1\,s against 1396.6\,s for the
60\,s fixed plan, a 57.4 per cent increase. Reducing only the maximum-green
multiplier from 2 to 1.25 lowers it to 409.2\,s, 70.7 per cent below that same
reference. Raising the multiplier to 3 instead produces 3846.3\,s. The sign
of this controller comparison therefore depends on its setting.

At all three loads, the lowest tested maximum-green multiplier has the lowest
mean within the presence-based sweep. The sweep does not bracket an optimum:
1.25 is its lower boundary. Appendix~\ref{app:results} gives all levels and
paired comparisons. In all 21 comparisons against the 60\,s reference, the ten
differences agree in sign. Each passes the campaign's Student-based Holm
correction at $p<0.001$; the corresponding Holm-adjusted sign-flip value is
approximately 0.041, conditional on its assumptions in Section~\ref{sec:stats}.

\subsection{Retiming the reference reduces the apparent benefit}

The 40\,s plan has the lowest mean of the three fixed settings at every tested
load. At 0.24 its endpoint is 473.1\,s, compared with 1396.6\,s at 60\,s and
3232.7\,s at 90\,s. Shortening the fixed plan improves the result without any
responsive control. Here too, the best tested setting lies at the sweep boundary.

Discharge-based termination has the lowest observed mean of all eight
configurations, but its advantage is much smaller against the retimed baseline.
Table~\ref{tab:retimed} reports the comparison selected against the best tested
fixed plan, using the unchanged family of 63 pairwise Student tests. All ten
paired differences favour termination at each load. These data compare the
specific implementations; they do not isolate discharge sensing from the
other differences between responsive and fixed-time control.

\begin{table}[htbp]
\centering\small
\caption{Discharge-based termination versus the 40\,s fixed plan. Negative
differences favour termination. Intervals are nominal paired 95 per cent
intervals; Student $p$ values are adjusted over 63 comparisons.}
\label{tab:retimed}
\begin{tabular}{@{}rrrrr@{}}
\toprule
Arrivals/s & Difference (s) & 95\% interval (s) & Reduction & Adj.\ $p$ \\
\midrule
0.12 & $-62.1$ & $[-64.9,\,-59.4]$ & 14.8\% & $<0.001$ \\
0.20 & $-70.1$ & $[-73.2,\,-66.9]$ & 15.8\% & $<0.001$ \\
0.24 & $-86.7$ & $[-94.6,\,-78.8]$ & 18.3\% & $<0.001$ \\
\bottomrule
\end{tabular}
\end{table}

\subsection{What the recorded traffic explains}

At the highest load, the default actuated configuration ends 96.8 per cent of
greens on absence of demand and 3.2 per cent at maximum green; mean green is
8.8\,s. Thus poor performance does not require most phases to reach their
maximum. These network-wide counts are not flow-weighted and do not identify
which movements constrain the network. They cannot rule out consequential
long greens at a small set of junctions.

The four lowest-mean configurations finish with approximately 88 to 108
vehicles on the road or waiting for entry, compared with 308 to 792 in the
other four. Removals range from about five per hour for the lower-mean group
to 249 for the highest-mean configuration. These are observed differences
in accumulated traffic and intervention under the common horizon. A removal
rule can keep occupancy bounded while journeys remain unserved, so these
counts do not demonstrate stability. Complete accounting and termination
summaries are in Appendix~\ref{app:results}.

\FloatBarrier
\section{Discussion and limitations}
\label{sec:threats}

\paragraph{Start-up loss and termination settings.}
Synthetic discharge checks in the repository find start-up lost time close to
zero, unlike the approximately two-second default used in traffic-engineering
manuals \citep{hcm7}. This discrepancy can favour frequent phase changes.
At 0.24 the discharge-based configuration averages 7.1\,s greens, close to the
7\,s minimum, and starts about 277 greens per signalised junction per hour.
The default actuated configuration starts about 246. Multiplying these counts
by an assumed start-up penalty would not correct the results: changed discharge
would also change queues and controller decisions. Explicit start-up-loss,
minimum-green and passage-timer sensitivity experiments are needed.
The 4\,s no-discharge rule could end green because of sparse arrivals or slow
start-up as well as downstream blockage. No direct blockage classification
was recorded, so its benefit cannot be assigned entirely to spillback avoidance.

\paragraph{Finite horizon and physical removals.}
The endpoint in Section~\ref{sec:tts} depends on inherited queues and on its
measurement boundaries. Equal warm-up durations do not establish comparable
stationarity. The retained occupancy-drift statistic treats serially correlated
samples as independent and is unsuitable as a stationarity test. Longer windows,
independent seeds and autocorrelation-aware diagnostics would be needed to
assess convergence. Removal and entry abandonment also require sensitivity
analysis: charging lost journeys through the window does not restore their
physical effects, and a vehicle stationary for 400\,s may be persistently
queued rather than irreversibly deadlocked.

\paragraph{One model and one demand pattern.}
The simulator has not been calibrated against local counts, trajectories or
signal plans. Its synthetic checks establish selected internal behaviours,
not real-network validity. The mostly two-phase network and demand spread across
directions may disadvantage long maximum greens relative to a strongly
directional peak with protected turns. Routes do not adapt to congestion;
there is no realistic turn-lane weaving, pedestrian phase demand, bus-stop dwell
or collision resolution. Shared assumptions do not protect relative rankings
from these omissions. Matched replication in an established simulator such as
SUMO \citep{lopez2018microscopic}, followed by field calibration, would be needed
for deployment claims.

\paragraph{Exploratory selection and uncertainty.}
The design followed an earlier comparison and did not reserve new seeds for
confirmation. Ten paired seeds provide limited information about the shape of
the difference distribution, and the Student-based comparison with the selected
40\,s plan is not independently confirmed by a family-adjusted sign-flip test.
Neither timing sweep brackets an optimum, and the discharge threshold was not
optimised. The reported reductions therefore describe tested settings under
this protocol, not the gains over optimally tuned fixed-time control.

\section{Conclusion}

On this simulated urban network, changing maximum green reverses the comparison
between presence-based actuation and the original fixed plan. A shorter fixed
plan substantially reduces the apparent advantage of responsive control, while
discharge-based termination retains a lower observed mean. The defensible
conclusion is methodological: compare timing settings and retimed references
before attributing a result to a controller family. Establishing a deployable
setting requires evidence about start-up loss, unresolved demand, other traffic
patterns and model validity beyond the present experiment.

\section*{Data and code availability}
\label{sec:repro}

Source is available under the PolyForm Noncommercial 1.0.0 licence at
\url{https://github.com/nitaiaharoni1/traffic-simulator}. The analysed campaign is
\texttt{ceiling-and-retiming}, saved at
\texttt{results/ceiling-and-retiming/2026-08-01T20-49-55-567Z}.
The accompanying \texttt{anc/ceiling-runs-summary.csv} contains all 240 run
summaries. Other ancillary files retain the broader campaign for transparency;
\texttt{anc/README.txt} distinguishes their scope. The exact experimental source
and runtime are not preserved in a complete versioned archive, which limits
long-term reproduction.

From the repository root, run \texttt{npx tsx scripts/paper-appendix.ts <results-dir>}
with that campaign directory to regenerate the full numerical tables and the
timing figure data. These commands analyse saved runs; they do not rerun traffic.
Build the manuscript with \texttt{make} inside \texttt{paper/}.

\section*{Acknowledgements}
This work used no external funding and no institutional resources.

\FloatBarrier
\bibliographystyle{plainnat}
\bibliography{refs}
\clearpage
\appendix

\section{Model parameters}
\label{app:params}

\begin{table}[h]
\centering
\small
\caption{Vehicle profiles. Length, minimum gap and emergency deceleration do not
vary between drivers; maximum acceleration and comfortable deceleration are scaled
by the driver's own sampled factors.}
\label{tab:vehicles}
\begin{tabular}{@{}lrrrrrrr@{}}
\toprule
Type & Share & Length & Min gap & $a_{\max}$ & $b_{\text{comf}}$ &
$b_{\text{emerg}}$ & $v_{\max}$ \\
 & & (m) & (m) & (m/s\textsuperscript{2}) & (m/s\textsuperscript{2}) &
(m/s\textsuperscript{2}) & (m/s) \\
\midrule
Car   & 85\% & 4.5  & 2.0 & 3.0 & 3.0 & 8.0 & 33.3 \\
Bus   & 8\%  & 12.0 & 3.0 & 1.5 & 2.5 & 6.0 & 22.2 \\
Truck & 7\%  & 16.0 & 4.0 & 1.0 & 2.0 & 5.0 & 25.0 \\
\bottomrule
\end{tabular}
\end{table}

\begin{table}[h]
\centering
\footnotesize
\caption{Other principal model and campaign parameters. Normal draws are shown
as mean and variance. Timing overrides are in Table~\ref{tab:design}.}
\label{tab:params}
\begin{tabular}{@{}p{2.6cm}p{5.2cm}p{6.6cm}@{}}
\toprule
Group & Parameter & Value \\
\midrule
\multirow{4}{=}{Driver sampling}
 & Desired speed factor & $\mathcal{N}(1.0,\,0.10^2)$ clamped $[0.5,\,1.6]$ \\
 & Desired time headway $T$ & $\mathcal{N}(1.5,\,0.30^2)$\,s clamped $[0.6,\,4]$ \\
 & Reaction time & $\mathcal{N}(1.0,\,0.20^2)$\,s clamped $[0.4,\,2]$ \\
 & Acceleration / deceleration factor &
   $\mathcal{N}(1.0,\,0.20^2)$, $\mathcal{N}(1.0,\,0.15^2)$ \\
\midrule
\multirow{3}{=}{Car following}
 & Free-road exponent & 4 \\
 & Look-ahead horizon & 200\,m, across links \\
 & Standstill rule & below 0.1\,m/s and offered $<0.15$\,m/s\textsuperscript{2} \\
\midrule
\multirow{4}{=}{Lane changing (MOBIL)}
 & Politeness & 0.5 \\
 & Advantage threshold & 0.2\,m/s\textsuperscript{2} \\
 & Safe braking for the new follower & 4.0\,m/s\textsuperscript{2} \\
 & Evaluation interval / cooldown & 0.1\,s / 4\,s \\
\midrule
\multirow{7}{=}{Signal plan synthesis}
 & Total green per cycle & 60\,s \\
 & Minimum green in the plan & 8\,s \\
 & Permitted (yielding) turns & enabled \\
 & Lane weight for a permitted-only claim & 0.35 \\
 & Yellow & $\mathrm{clamp}(1 + v/(2\cdot 3),\,3,\,6)$\,s \\
 & All-red & $\mathrm{clamp}((L_{\text{move}} + 12)/8,\,1,\,6)$\,s \\
 & Junction clustering radius & 40\,m, up to 6 nodes \\
\midrule
\multirow{4}{=}{Control skeleton}
 & Minimum green & 7\,s \\
 & Maximum green & $2\times$ the plan's green for that phase \\
 & Passage timer & none; terminate on first clear scan after minimum green \\
 & No-discharge threshold (\strat{actuated-truncating}) & 4\,s, checked after minimum green \\
\midrule
\multirow{2}{=}{Detection}
 & Detector update & 1\,Hz \\
 & Controller scan cadence & 0.1\,s \\
\midrule
\multirow{4}{=}{Gap acceptance}
 & Critical gap: left / straight / right / u-turn & 4.0 / 3.0 / 2.5 / 5.0\,s \\
 & Look-ahead horizon & 8\,s \\
 & Patience limit & 12\,s \\
 & Critical gap floor & 1.5\,s \\
\midrule
\multirow{4}{=}{Demand}
 & Trip mix (internal / in / out / through) & 55 / 20 / 20 / 5\,\% \\
 & Inter-arrival distribution & exponential \\
 & Entry abandonment & 300\,s \\
 & Origin placement & uniform along the origin lane (interior trips) \\
\midrule
\multirow{2}{=}{Measurement}
 & Stopped-speed threshold & 0.5\,m/s \\
 & Deadlock removal & stationary for 400\,s \\
\midrule
\multirow{3}{=}{Integration}
 & Timestep & 0.1\,s \\
 & Warm-up / measured window & 32{,}400\,s / 3{,}600\,s \\
 & Seeds per cell & 10 (integers 1 to 10) \\
\bottomrule
\end{tabular}
\end{table}

\clearpage

\section{Full results}
\label{app:results}

All 8 configurations at 3 loads, over 10 paired seeds. Tables are generated from the saved campaign by \texttt{scripts/paper-appendix.ts}. Population counts include road vehicles and entry waiting. Removed and abandoned count physical removals and entry-queue losses, respectively. The paired tables compare against \strat{fixed-time}. Their intervals are nominal 95 per cent Student intervals, not simultaneous intervals; raw and adjusted $p$ values use the paired Student test. Bold configurations pass Holm correction. The last column gives unadjusted exact sign-flip values, with the assumptions and resolution limits in Section~\ref{sec:stats}.

\begin{table}[htbp]
\centering
\footnotesize
\setlength{\tabcolsep}{3.2pt}
\caption{Mean endpoint and traffic accounting at 0.12 arrivals per second.}
\label{tab:all012}
\begin{tabular}{@{}lrrrrrr@{}}
\toprule
Strategy & Time spent & Initial & Final & Completed & Removed & Abandoned \\
 & (s) & population & population & (per h) & (per h) & (per h) \\
\midrule
\strat{fixed-time} & 458.6 & 51.5 & 56.0 & 422.1 & 1.4 & 3.0 \\
\strat{fixed-green40} & 420.8 & 47.7 & 51.1 & 423.2 & 1.4 & 3.0 \\
\strat{fixed-green90} & 565.7 & 62.7 & 69.8 & 419.4 & 1.4 & 3.1 \\
\strat{actuated} & 364.1 & 41.0 & 44.1 & 423.5 & 1.4 & 3.0 \\
\strat{actuated-max125} & 360.8 & 39.9 & 43.8 & 422.7 & 1.4 & 3.0 \\
\strat{actuated-max150} & 361.4 & 39.9 & 44.2 & 422.3 & 1.4 & 3.0 \\
\strat{actuated-max300} & 371.4 & 40.7 & 45.6 & 421.7 & 1.4 & 3.0 \\
\strat{actuated-truncating} & 358.7 & 38.7 & 42.7 & 422.6 & 1.4 & 3.0 \\
\bottomrule
\end{tabular}
\end{table}

\begin{table}[htbp]
\centering
\footnotesize
\setlength{\tabcolsep}{3.2pt}
\caption{Mean endpoint and traffic accounting at 0.2 arrivals per second.}
\label{tab:all02}
\begin{tabular}{@{}lrrrrrr@{}}
\toprule
Strategy & Time spent & Initial & Final & Completed & Removed & Abandoned \\
 & (s) & population & population & (per h) & (per h) & (per h) \\
\midrule
\strat{fixed-time} & 803.1 & 150.8 & 150.4 & 700.3 & 8.5 & 6.9 \\
\strat{fixed-green40} & 442.3 & 82.6 & 81.5 & 708.4 & 4.0 & 4.0 \\
\strat{fixed-green90} & 2024.1 & 360.7 & 363.6 & 631.2 & 66.1 & 15.1 \\
\strat{actuated} & 467.6 & 83.2 & 87.1 & 702.9 & 4.1 & 4.4 \\
\strat{actuated-max125} & 380.9 & 68.3 & 68.6 & 707.1 & 4.0 & 3.9 \\
\strat{actuated-max150} & 388.4 & 70.4 & 70.0 & 707.6 & 4.0 & 4.1 \\
\strat{actuated-max300} & 2609.2 & 431.0 & 460.6 & 542.1 & 121.8 & 21.8 \\
\strat{actuated-truncating} & 372.2 & 67.0 & 66.5 & 707.9 & 4.0 & 3.9 \\
\bottomrule
\end{tabular}
\end{table}

\begin{table}[htbp]
\centering
\footnotesize
\setlength{\tabcolsep}{3.2pt}
\caption{Mean endpoint and traffic accounting at 0.24 arrivals per second.}
\label{tab:all024}
\label{tab:all}
\begin{tabular}{@{}lrrrrrr@{}}
\toprule
Strategy & Time spent & Initial & Final & Completed & Removed & Abandoned \\
 & (s) & population & population & (per h) & (per h) & (per h) \\
\midrule
\strat{fixed-time} & 1396.6 & 304.6 & 307.6 & 808.3 & 49.3 & 15.6 \\
\strat{fixed-green40} & 473.1 & 110.5 & 107.5 & 869.6 & 4.8 & 4.8 \\
\strat{fixed-green90} & 3232.7 & 655.1 & 686.4 & 618.2 & 186.4 & 40.3 \\
\strat{actuated} & 2198.1 & 447.1 & 471.2 & 701.7 & 124.5 & 25.9 \\
\strat{actuated-max125} & 409.2 & 91.6 & 93.1 & 864.8 & 4.9 & 5.0 \\
\strat{actuated-max150} & 430.6 & 95.0 & 96.9 & 864.4 & 4.8 & 5.1 \\
\strat{actuated-max300} & 3846.3 & 765.9 & 792.4 & 549.8 & 249.3 & 50.6 \\
\strat{actuated-truncating} & 386.4 & 88.9 & 87.6 & 867.9 & 4.8 & 4.8 \\
\bottomrule
\end{tabular}
\end{table}

\begin{table}[htbp]
\centering
\footnotesize
\setlength{\tabcolsep}{3.6pt}
\caption{Paired endpoint differences at 0.12 arrivals per second. Holm adjustment retains the campaign's family of 21 comparisons.}
\label{tab:paired012}
\begin{tabular}{@{}lrrrrr@{}}
\toprule
Strategy & Difference (s) & 95\% CI & Raw $p$ & Adj.\ $p$ & Perm.\ $p$ \\
\midrule
\textbf{\strat{actuated-truncating}} & $-$99.8 ($-$21.8\%) & $[-108,\,-92]$ & $<$0.001 & $<$0.001 & 0.002 \\
\textbf{\strat{actuated-max125}} & $-$97.8 ($-$21.3\%) & $[-106,\,-90]$ & $<$0.001 & $<$0.001 & 0.002 \\
\textbf{\strat{actuated-max150}} & $-$97.1 ($-$21.2\%) & $[-105,\,-89]$ & $<$0.001 & $<$0.001 & 0.002 \\
\textbf{\strat{actuated}} & $-$94.5 ($-$20.6\%) & $[-103,\,-86]$ & $<$0.001 & $<$0.001 & 0.002 \\
\textbf{\strat{actuated-max300}} & $-$87.2 ($-$19.0\%) & $[-95,\,-79]$ & $<$0.001 & $<$0.001 & 0.002 \\
\textbf{\strat{fixed-green40}} & $-$37.7 ($-$8.2\%) & $[-45,\,-31]$ & $<$0.001 & $<$0.001 & 0.002 \\
\textbf{\strat{fixed-green90}} & $+$107.1 ($+$23.4\%) & $[73,\,142]$ & $<$0.001 & $<$0.001 & 0.002 \\
\bottomrule
\end{tabular}
\end{table}

\begin{table}[htbp]
\centering
\footnotesize
\setlength{\tabcolsep}{3.6pt}
\caption{Paired endpoint differences at 0.2 arrivals per second. Holm adjustment retains the campaign's family of 21 comparisons.}
\label{tab:paired02}
\begin{tabular}{@{}lrrrrr@{}}
\toprule
Strategy & Difference (s) & 95\% CI & Raw $p$ & Adj.\ $p$ & Perm.\ $p$ \\
\midrule
\textbf{\strat{actuated-truncating}} & $-$430.9 ($-$53.7\%) & $[-506,\,-355]$ & $<$0.001 & $<$0.001 & 0.002 \\
\textbf{\strat{actuated-max125}} & $-$422.2 ($-$52.6\%) & $[-497,\,-347]$ & $<$0.001 & $<$0.001 & 0.002 \\
\textbf{\strat{actuated-max150}} & $-$414.7 ($-$51.6\%) & $[-490,\,-340]$ & $<$0.001 & $<$0.001 & 0.002 \\
\textbf{\strat{fixed-green40}} & $-$360.8 ($-$44.9\%) & $[-435,\,-287]$ & $<$0.001 & $<$0.001 & 0.002 \\
\textbf{\strat{actuated}} & $-$335.5 ($-$41.8\%) & $[-418,\,-253]$ & $<$0.001 & $<$0.001 & 0.002 \\
\textbf{\strat{fixed-green90}} & $+$1221.0 ($+$152.0\%) & $[1105,\,1337]$ & $<$0.001 & $<$0.001 & 0.002 \\
\textbf{\strat{actuated-max300}} & $+$1806.1 ($+$224.9\%) & $[1613,\,2000]$ & $<$0.001 & $<$0.001 & 0.002 \\
\bottomrule
\end{tabular}
\end{table}

\begin{table}[htbp]
\centering
\footnotesize
\setlength{\tabcolsep}{3.6pt}
\caption{Paired endpoint differences at 0.24 arrivals per second. Holm adjustment retains the campaign's family of 21 comparisons.}
\label{tab:paired024}
\label{tab:paired}
\begin{tabular}{@{}lrrrrr@{}}
\toprule
Strategy & Difference (s) & 95\% CI & Raw $p$ & Adj.\ $p$ & Perm.\ $p$ \\
\midrule
\textbf{\strat{actuated-truncating}} & $-$1010.2 ($-$72.3\%) & $[-1051,\,-969]$ & $<$0.001 & $<$0.001 & 0.002 \\
\textbf{\strat{actuated-max125}} & $-$987.4 ($-$70.7\%) & $[-1027,\,-948]$ & $<$0.001 & $<$0.001 & 0.002 \\
\textbf{\strat{actuated-max150}} & $-$965.9 ($-$69.2\%) & $[-1008,\,-924]$ & $<$0.001 & $<$0.001 & 0.002 \\
\textbf{\strat{fixed-green40}} & $-$923.4 ($-$66.1\%) & $[-968,\,-879]$ & $<$0.001 & $<$0.001 & 0.002 \\
\textbf{\strat{actuated}} & $+$801.5 ($+$57.4\%) & $[587,\,1016]$ & $<$0.001 & $<$0.001 & 0.002 \\
\textbf{\strat{fixed-green90}} & $+$1836.1 ($+$131.5\%) & $[1621,\,2051]$ & $<$0.001 & $<$0.001 & 0.002 \\
\textbf{\strat{actuated-max300}} & $+$2449.7 ($+$175.4\%) & $[2225,\,2674]$ & $<$0.001 & $<$0.001 & 0.002 \\
\bottomrule
\end{tabular}
\end{table}

\begin{table}[htbp]
\centering
\footnotesize
\setlength{\tabcolsep}{4pt}
\caption{Phase-controller telemetry, averaged across seeds. Ceiling is the percentage of recorded greens reaching the configuration's maximum green. Fixed-time arms do not record these counters and are omitted.}
\label{tab:mechanism-full}
\begin{tabular}{@{}l rrrrrr@{}}
\toprule
& \multicolumn{3}{c}{ceiling (\%)} & \multicolumn{3}{c}{mean green (s)} \\
\cmidrule(lr){2-4}\cmidrule(l){5-7}
Strategy & 0.12 & 0.2 & 0.24 & 0.12 & 0.2 & 0.24 \\
\midrule
\strat{actuated} & 0.1 & 0.5 & 3.2 & 7.2 & 7.4 & 8.8 \\
\strat{actuated-max125} & 0.1 & 0.2 & 0.4 & 7.2 & 7.3 & 7.3 \\
\strat{actuated-max150} & 0.1 & 0.3 & 0.5 & 7.2 & 7.3 & 7.4 \\
\strat{actuated-max300} & 0.1 & 2.3 & 4.1 & 7.2 & 8.9 & 10.4 \\
\strat{actuated-truncating} & 0.0 & 0.0 & 0.0 & 7.1 & 7.1 & 7.1 \\
\bottomrule
\end{tabular}
\end{table}

\clearpage
\section{Recorded simulator checks}
\label{app:checks}

Table~\ref{tab:checks} summarises the evidence available to judge the simulation
instrument. These are previously recorded checks, not new runs or additional
replications of the timing experiment. The source descriptions and scenario
settings are retained in \texttt{experiments/README.md}; the timestep campaign is
\texttt{results/timestep-convergence/2026-07-30T17-34-06-786Z}.
None of these checks compares the study network with field observations.

\begin{table}[htbp]
\centering\small
\caption{Selected internal checks and model discrepancies. Synthetic scenarios
share model assumptions with the main experiment, so agreement is evidence of
internal consistency, not independent validation of field performance.}
\label{tab:checks}
\begin{tabular}{@{}p{3.2cm}p{11.4cm}@{}}
\toprule
Check and setting & Recorded evidence and interpretation \\
\midrule
Baseline against itself, synthetic grid &
Eight paired seeds gave exactly zero differences on all metrics. This checks
whether the comparison procedure invents a difference. \\[4pt]
Timestep sensitivity, synthetic grid &
At 0.1 arrivals/s, comparing 0.1\,s with $1/60$\,s found no detectable
difference in delay, throughput or mean speed; reported paired uncertainty
bounds were 1.6, 0.6 and 1.0 per cent, respectively. The 80 runs had no removals, denied entries or
frozen vehicles. This evidence applies to those scenarios, not every regime. \\[4pt]
Saturation discharge, isolated approach &
Ten seeds gave a car-only headway of 2.369\,s (between-seed standard deviation
0.076\,s), or 1,520 vehicles/h/lane. The default fleet gave 1,259.
The Highway Capacity Manual reference is 1,900 \citep{hcm7}; the discrepancy
is not evidence that errors on the study network are conservative. \\[4pt]
Start-up lost time, isolated approach &
Start-up lost time was close to zero, versus a manual default near 2\,s
\citep{hcm7}. This is a material discrepancy for short greens. Total lost time,
including clearance effects, was 4.88\,s per phase; total and start-up loss
must not be conflated. \\[4pt]
Signal delay, isolated crossroads &
With three seeds and straight-through Poisson demand, simulated signal delay
was within 10 per cent of Webster's uniform term plus the incremental term
for degrees of saturation 0.3 to 0.9, after subtracting the model's own
no-signal residual. Capacity in that comparison was measured in the same
model, not independently observed. \\[4pt]
Vehicle overlap, moving city extract &
Overlaps deeper than 0.5\,m per 1,000 samples were 89.2, 0.0, 1.7 and 107.5
at timesteps $1/60$, 0.05, 0.1 and 0.2\,s. The relation was non-monotone:
a finer step did not guarantee fewer overlaps. Car following does not enforce
a hard collision-free constraint. \\[4pt]
Measurement instrumentation, study district &
Enabling per-junction measurement produced a bit-identical vehicle trace to
an uninstrumented run, alongside 177,363 junction rows. This checks that
measurement does not perturb the model; it does not validate the traffic model. \\
\bottomrule
\end{tabular}
\end{table}

\end{document}